\documentclass[pre,aps,twocolumn,showpacs,floatfix,longbibliography]{revtex4-1}
\usepackage{graphicx}
\usepackage{epsf}
\usepackage{amsmath}
\usepackage{amssymb}
\usepackage{xcolor}
\usepackage{mdframed}
\usepackage{color}

\begin{document}

\title{Statistical Theory of Selectivity and Conductivity in Biological Channels}

\author{D.~G.~Luchinsky$^{1,2}$, W.A.T.~Gibby$^1$, I. Kaufman$^1$, D.A. Timucin$^3$, P.V.E.~McClintock$^1$}

\affiliation{$^1$Department of Physics, Lancaster University, Lancaster LA1 4YB, UK.}
\email{dmitry.g.luchinsky@nasa.gov}%
\email{w.gibby@lancaster.ac.uk}%
\email{i.kaoufman@lancaster.ac.uk}%
\email{p.v.e.mcclintock@lancaster.ac.uk}%

\vspace{0.4cm}

\affiliation{$^2$SGT Inc., Greenbelt, MD, 20770, USA}

\affiliation{$^3$ARC, Moffett Field, CA, 94035, USA}
\email{dogan.a.timucin@nasa.gov}%


%
\date{\today}
\begin{abstract}
We present an equilibrium statistical-mechanical theory of selectivity in biological ion channels. In doing so, we introduce a grand canonical ensemble for ions in a channel's selectivity filter coupled to internal and external bath solutions for a mixture of ions at arbitrary concentrations, we use linear response theory to find the current through the filter for small gradients of electrochemical potential, and we show that the conductivity of the filter is given by the generalized Einstein relation. We apply the theory to the permeation of ions through the potassium selectivity filter, and are thereby able to resolve the long-standing paradox of why the high selectivity of the filter brings no associated delay in permeation. We show that the Eisenman selectivity relation follows directly from the condition of diffusion-limited conductivity through the filter. We also discuss the effect of wall fluctuations on the filter conductivity.
\end{abstract}



\maketitle



\section{Introduction}\label{s:introduction}

The statistical mechanical and kinetic theory of multiple ions competing for binding sites has enjoyed renewed interest in recent years because of a diversity of important applications, including e.g.\ surface adsorption~\cite{Romanielo2015} and adsorption in nanopores~\cite{Roque-Malherbe2007a}, charge transport in porous electrodes~\cite{Bisquert2008,Bazant2013}, selective permeability of ion transport proteins~\cite{Bostick2009}, and nanobiology~\cite{Jain2015,Feng2016}.

In this context it is instructive to revisit, from the standpoint of statistical physics, the long-standing problem of the conductivity and selectivity of biological ion channels~\cite{Eisenman61,Mullins1959,MacKinnon2001b,MacKinnon2003,Noskov2007a,Dixit2009a,Roux:11,Piasta:2011,Dixit2011a,Roux:2014}: how can channels select K$^+$ ions at a ratio of 1000:1 over the (smaller) Na$^+$ ions, and nonetheless still conduct the K$^+$ ions at almost the diffusion-limited rate, approaching $10^8$ ions per second?

The unusual properties of these channels are mainly determined by the structure of their nano-scale selectivity filters.

The canonical K$^+$ selectivity filter is formed by a narrow $\sim12$\,\AA~long tunnel. It is lined by 20 oxygen atoms comprising four K$^+$-binding sites, numbered S$_1$ -- S$_4$ from the extracellular to the intracellular side~\cite{Piasta:2011,Zhou2001,Nimigean2011b}. An additional binding site S0 located at the extracellular mouth of the filter is partly hydrated and partly coordinated by the carbonyl oxygen.

Following the original ideas of~\cite{Mullins1959,Eisenman61} a thermodynamic analysis of this structure attributes selectivity mainly~\cite{Eisenman61,Bezanilla1972,Eisenman:83,Noskov2007a,Dixit2009a,Roux:11,Dixit2011a,Roux:2014} to the difference in the free energy barrier for K$^+$ and Na$^+$ ions to enter the pore. This difference is usually expressed in terms of the excess chemical potential as
\begin{equation}\label{eq:eisenman}
	\varDelta\varDelta\bar{\mu}_{K,Na} = (\bar{\mu}^{c}_{Na}-\bar{\mu}_{Na})-(\bar{\mu}^{c}_{K}-\bar{\mu}_{K}),
\end{equation}
where $\varDelta\bar{\mu}_{i} = \bar{\mu}_{i}-\bar{\mu}^{c}_{i}$ is the difference in excess chemical potential for an ion of type $i$ in the bulk and in the channel.

It was observed, both in experiments and in molecular dynamics simulations, that the four binding sites are populated by K$^+$ ions with approximately the same probability ~\cite{MacKinnon2001b,MacKinnon2003}. Each pair of ions is usually separated by at least one intervening water molecule, and ``knock-on'' like conduction occurs by transitions of the filter between two states where it contains either two or three K$^+$ ions respectively.

The dominant view about the selectivity and conductivity of this filter is based on the idea of a ``snug-fit'' ~\cite{Mullins1959,Bezanilla1972,Doyle:98a} into a rigid filter that provides an isoenergetic, aquomimetic diffusion pathway for K$^+$, whereas a Na$^+$ ion is confronted by large free energy barrier~\cite{Eisenman61,Piasta:2011}.

However, this simple and elegant picture does not agree well with the observed flexibility of the channel walls~\cite{Allen2004a} or the experimentally observed multi-ion nature of the transition mechanism~\cite{MacKinnon2001b}. It was proposed~\cite{MacKinnon2003} that, in order to explain the observed phenomena, one may need to develop a non-equilibrium theory~\cite{Roux:2014,Derebe2011b} and/or take account of the multi-ion, multi-binding-site, nature of the conduction mechanism.

In what follows, however, we address the problem by application of equilibrium statistical mechanics. The statistical properties of the filter, coupled to solutions with a mixture of conducting ions, are obtained by introducing the grand canonical ensemble for ions within the filter. The permeating current is considered in the linear response regime. The interaction of charges within the filter is taken into account using the electrostatic self-energy approximation.

We apply our approach to the analysis of ion permeation through the potassium channel. We show that the selectivity {\it vs.} conductivity paradox can be resolved because the equilibrium selectivity equation follows directly from the condition for barrier-less conduction.

In our derivation we follow the spirit and ideas of our earlier work \cite{Kaufman:13a,Kaufman:12,Kaufman:15}. For a single type of conducting ion, we recover the distribution of ions in the filter obtained previously ~\cite{Roux1999,Roux:04}.

The paper is organized as follows. In the next section we introduce a statistical mechanical theory of a generic selectivity filter with multiple binding sites coupled to solutions with mixed ions. In Sec.~\ref{s:conduction} we discuss the conductivity of the filter in the linear response regime. In Sec.~\ref{s:potassium_filter} we apply the theory to the analysis of conductivity and selectivity of the potassium channel. Finally, in Sec.~\ref{s:conclusions}, we summarise the results obtained and suggest some future directions of the research.

\section{ Model of the selectivity filter}\label{s:model_of_SFF}

\begin{figure}[h!]
	\centering
	\includegraphics[width = 8cm, height = 4.5cm]{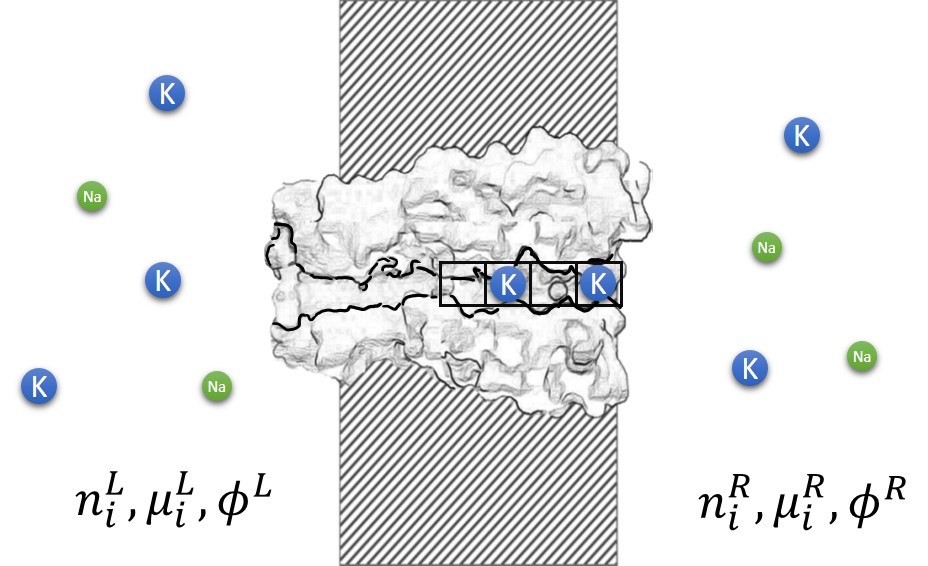}
	\caption{Schematic representation of the selectivity filter coupled to bath solutions on the left and right. The binding sites S$_1$ -- S$_4$ for cations are indicated by black squares. In this instance, two of them happen to be occupied by K$^+$ ions, shown as small blue circles.}
	\label{fig:channel}
\end{figure}

We consider an ion channel diffusively and thermally coupled to two bath solutions that may contain $m$ different types of conducting ion, with $n_i$ being the number of ions of $i$-th type. In a typical situation of interest there are several binding sites in the channel pathway that can be occupied either by water molecules or by ions.  The channel is shown schematically in Fig.~\ref{fig:channel}

In a single-file selectivity filter, each binding site can accommodate only one ion.

\subsection{State space}\label{ss:state_space}

Let us consider first an example where there are $K$ distinguishable binding sites, i.e.\ such that the binding energy of each of them is different.

In a solution with $ m $ different types of permeable ion, the state space of the selectivity filter is defined by $K$ numbers corresponding to the type of ion at each binding site
\begin{equation}\label{eq:state_space_D}
\left\lbrace j_1, j_2,\ldots, j_K \right\rbrace .
\end{equation}
For an empty site, $j_i = 0$. In general, the number of energy levels in such a system is $ (m +1) ^K $, where the 1 has been added to account for water molecules in the filter. Some of these energy levels are degenerate, because we may exchange the positions of ions of the same type without changing the energy. Additional restrictions may apply to reduce the number of different energy levels. For example, based on experimental evidence, one might require that there be at least one water molecule separating the cations in the filter.

For indistinguishable binding sites (i.e.\ sites having the same binding energy for a given type of ions) the state space is reduced and can be characterized by the number of ions of each type $n_j$ that are currently inside the filter
\begin{equation}\label{eq:state_space_I}
\left\lbrace n_j \right\rbrace = \left\lbrace n_1, n_2,\ldots, n_m \right\rbrace ,
\end{equation}
where $\sum_i^{m} n_i \le K$. Taking into account that indistinguishable binding sites can be filled either with water molecules, or with one ion out of the $ m $ types of ion that we consider, there are $(K+m)!/m!/K! $ different energy levels in the system. This total is, however, subject to the possible additional constraints mentioned above.

In what follows we will assume for simplicity that the filter can be represented by a set of indistinguishable binding sites; we will leave the more general case of distinguishable binding sites for future publications.

To find binding probabilities we will derive a grand canonical ensemble for ions in the filter coupled to the mixed solutions. First, we find the total energy of system, counting explicitly the number of ions in each solution and in the filter. Next, we separate the degrees of freedom for the solutions and filter, which allows us to introduce an effective grand canonical ensemble for the filter.

\subsection{Energy of the system}\label{ss:energy}

The electrochemical potential $\left( \mu^s_i = \left( \partial E^s/\partial n_i^s\right)_{S,V}\right) $ of the  $ i$-th type of ion in the bulk solutions in the applied field can be written as
\begin{eqnarray}\label{eq:chem-pot-b}
	\mu^s_i = kT\ln\left( x^s_i\right) + \bar{\mu}_i + q\phi^s,  \qquad s  = L, R.
\end{eqnarray}
Here $x^s_i = \frac{n^s_i}{N^s}$ is the number of moles, $\bar{\mu}^s_i$ is the excess chemical potential, $\phi$ is the electric scalar potential, $n_i^s$ is the number of ions and $N^s = n_w^s + \sum_i^m n_i^s$ is the total number of conducting ions and water molecules in the left $s = L$ or right $ s=R$ solution, as illustrated in Fig.~\ref{fig:channel}.

Note that in (\ref{eq:chem-pot-b}) we have neglected interactions between ions of the same polarity, which is a reasonable approximation for bulk solutions under physiological conditions.

The ions in the channel are characterized by their excess chemical potential $ \bar{\mu} ^c_i$, static electric potential $ \phi^c $ (i.e.\ within the potential drop between the entrance and exit of the selectivity filter), and the energy of their electrostatic interaction with the channel charge which we model as being on its walls $\varepsilon\left( \{n_j\}, n_f\right)$.

Note, that in general one cannot say how many ions entered the channel from the left ($n'_i$) or right ($n''_i$) solutions. So we assume that the number of ions in the filter is $n_i = n'_i + n''_i$. Similarly, for the water molecules, we have $n_w  = n'_w  + n''_w $.

For a specific state with the set of $\{n_1, n_2,\dots, n_m\}$ ions and $ n _w $ water molecules in the filter, the energy of the system can be written
\begin{eqnarray}\label{eq:excited_state_energy}
&&\hspace{-0.25cm}E\left( \left\lbrace n_j \right\rbrace,n_f\right) = E_0 + (n^L_w-n'_w)\mu^L_w + (n^R_w-n''_w)\mu^R_w +  \nonumber\\
&&~\sum_i( n^L_i-n'_i) \mu^L_i + \sum_i( n^R_i-n''_i) \mu^R_i+n_w\mu_w^s+ \\
&&\sum_i n_i(\bar{\mu}_i^c+q\phi^c)  + kT\ln n_w!\prod_i n_i!  +   \varepsilon\left( \{n_j\}, n_f\right) . \nonumber
\end{eqnarray}
In equilibrium the following condition holds separately for each of the permeating species
\[\mu^L_w =\mu^R_w, \qquad  \mu^L_i =\mu^R_i,\]
and we can rewrite equation (\ref{eq:excited_state_energy}) in terms of either the left or right chemical potentials
\begin{eqnarray}\label{eq:excited_state_energy_s}
&&E\left( \left\lbrace n_j \right\rbrace,n_f\right) = E_0 + (N_w-n_w)\mu^s_w +  \nonumber\\
&&\quad\sum_i(N_i-n_i) \mu^s_i + n_w\bar{\mu}_w^c + kT\ln n_w! +  \\
&&\sum_i n_i(\bar{\mu}_i^c+q\phi^c)  + kT\ln\prod_i n_i!  +   \varepsilon\left( \{n_j\}, n_f\right), \nonumber
\end{eqnarray}
where $s$ is either $L$ or $R$.

The energy levels with non-zero numbers of ions in the filter can be considered as excitations of the ground state corresponding to an empty filter filled with water molecules.

In (\ref{eq:excited_state_energy_s}) $E_0 = TS-pV$ is the thermodynamic part of the energy of the bulk solution. Next there are two terms related to the energy required to add $(N_w-n_w)$ water molecules and $\sum(N_i-n_i)$ ions to the bulk solution. $N_w$ is the total number of water molecules with excess chemical potential $ \mu_w $ in the bath of volume $V$, and $N_i $ is the total number of ions of the $i^{\rm th}$ type in the system.

The term $\sum_i (N_i-n_i)\mu^s_i$ takes explicit account of the fact that the number of ions in the solution is changed due to their transfer to the filter. However, we neglect the corresponding changes in chemical potential $\mu_i^s$ and mole fraction $x_i^s$ in the solution because they are very small. The latter assumption corresponds to the approximation $(N_i -n_i)!/n_i! \approx (N_i)^{n_i}$ \cite{Roux1999}.

The last two terms in the second row correspond to the energy required to insert $n_w$ water molecules into the filter. The term $\left( kT\ln n_w!\right)$ takes into account permutation of the water molecules in the filter.

The last row describes the energy required to add $n_i$ ions of $i^{\rm th}$ type to the filter. The term  $\left( \prod_i n_i!\right)$ takes into account the permutations of each type of ion in the filter with the indistinguishable binding sites.

The last term in (\ref{eq:excited_state_energy_s}) plays an important role in ionic conduction and selectivity, and it needs to be considered in more detail.

\subsection{Ions-filter interaction }\label{ss:ion_filter_interaction}

The final term in equation (\ref{eq:excited_state_energy}) corresponds to the ion-ion and ion-filter interactions inside the filter. It is known that this interaction is important and cannot be neglected.

In general, the estimation of this interaction is a nontrivial problem \cite{Shklovskii:05,Kharkyanen:10}. However, in the first approximation it can be accounted for as the electrostatic self-energy of the ions in the filter~\cite{Shklovskii:05,Kaufman:13a}. The latter can be written as
\begin{equation}\label{eq:electrostatic_energy}
\varepsilon(\{n_i\}, n_f) = \frac{q^2}{2C}\left(\sum_i n_i+n_f \right)^2 ,
\end{equation}
where $\sum_i n_i$ is the total number of ions in the channel and $n_f$ is the fractional number of fixed unit charges $q$ on the channel wall. We note that $n_f$ is one of the main mutation parameters in the system and that it usually takes negative real values between 0 and $-6$.

The energy defined by equation (\ref{eq:electrostatic_energy}) is equivalent to the electrostatic energy of the electrons in a quantum dot \cite{Beenakker:91,Alhassid:00} and, as in the case of a quantum dot, $C$ can be identified~\cite{Shklovskii:05} as the channel capacitance.

Assuming that the electric field in the filter does not penetrate the protein walls, the channel capacitance $C$ can be estimated as
\begin{equation}\label{eq:capacitance}
C\approx\frac{4\pi\epsilon_0\epsilon_wR^2}{L}.
\end{equation}
For a more accurate evaluation of $C$ see e.g.\ \cite{Finkelstein2006a}.

Qualitatively, the results obtained do not depend on the specific form of the interaction, which can be written as a general quadratic form $\sum_{ik}\beta_{ik}n_i n_k/2N$~\cite{Landau:80}, where the sum runs over all the charges, including the wall charge; see also ~\cite{Kitzing1992} for an example of a different type of interaction.

\subsection{Grand canonical ensemble of the selectivity filter}\label{ss:GCE}

To introduce the grand canonical ensemble for ions within the filter one has to separate its degrees of freedom from those of the ions in the bulk solution. This can be done by simple rearrangement of the terms in equation (\ref{eq:excited_state_energy_s}), which gives
\begin{eqnarray}\label{eq:filter_state_energy}
	&&E\left( \left\lbrace n_j \right\rbrace,n_f\right) = E_0 +  N_w\mu^s_w + \sum_i  N_i\mu^s_i + \nonumber\\
	&&\quad - n_w\Delta\bar{\mu}_w^s + kT\ln n_w! - kT \sum_i n_i ln x^s_i +  \\
	&&- \sum_i n_i\Delta\tilde{\mu}_i^s + kT\ln\prod_i n_i! + \varepsilon\left( \{n_j\}, n_f\right). \nonumber
\end{eqnarray}
Here $\Delta\tilde{\mu}_i^s = \Delta\bar{\mu}_i ^s + q\Delta\phi^s$ is the difference in excess chemical potential and electrostatic potential of the ions in the bulk and in the channel. $\Delta\bar{\mu}_i^s$ is defined in the Eisenman equation (\ref{eq:eisenman}) and $ \Delta \phi^s = (\phi^s - \phi^c)$ is the electrostatic potential drop between the $s^{\rm th}$ solution and the filter.

For water molecules we introduce similar notation $\Delta\bar{\mu}_w^s  = \bar{\mu}^s_w - \bar{\mu}_w^c$ to emphasize that they can be treated on equal grounds with the conducting ions. In what follows, however, we will neglect for simplicity the contribution of the water molecules to the filter energy.

We note that the first three terms in equation (\ref{eq:filter_state_energy}) correspond to the constant energy of the bulk solutions, while the remaining terms depend on the filter's degrees of freedom. This allows us to factorize the partition function for the whole system into a product of two terms related respectively to the solution and to the filter.

Once the degrees of freedom have been separated, we can write the grant canonical ensemble for the filter \cite{Baierlein:99,Baierlein2001} in a form
\begin{equation}\label{eq:grand_canonical}
	P(\left\lbrace n_j \right\rbrace,n_f) = \mathcal{Z}^{-1}\prod_{i=1}^m\frac{(x_i^s)^{n_i}}{n_i!} e^{\frac{\sum_i n_i\Delta\tilde{\mu}_i^s-\varepsilon\left( \left\lbrace n_j \right\rbrace,n_f\right) }{kT}},
\end{equation}
that depends on the parameter $n_f$. The resultant grand partition function is
\begin{equation}\label{eq:PF_grand}
\mathcal{Z} = \sum_{\substack{\left\lbrace n_j \right\rbrace \\ \sum n_j \le K}} \prod_{i=1}^m\frac{(x_i^s)^{n_i}}{n_i!} e^{\frac{\sum_i n_i\Delta\tilde{\mu}_i-\varepsilon\left( \left\lbrace n_j \right\rbrace,n_f\right) }{kT}},
\end{equation}
where $\left\lbrace n_j  \right\rbrace $ runs though all possible configurations of conducting ions in the filter.

The grand potential of the filter is then given by
\begin{eqnarray}\label{eq:grand_potential}
\Omega = -kT \ln\mathcal{Z}.
\end{eqnarray}

\subsection{Operational definition of $\mu^c$}\label{ss:operational_operational}

To enable a comparison of the value $\mu^c$ with the results of molecular dynamics simulations, we note that equation (\ref{eq:PF_grand}) can be related to the effective Grand Canonical Partition function introduced in~\cite{Roux1999,Roux:04}.

This can be done if we consider only one type of conducting ion and note that the configuration integral defined in equation (22) of~\cite{Roux1999} provides estimates of the excess chemical potential and of the energy of electrostatic interaction of $n$ ions in the channel
\begin{equation}\label{eq:config_integr}
e^{-\beta\left(  n\bar{\mu}^c  +  \varepsilon\left( n ,n_f\right)\right) } \approx \int_cd\bf{r}_1\ldots\int_cd\bf{r}_ne^{-\beta \mathcal{W}(\bf{r}_1...\bf{r}_n)}.
\end{equation}
$ \mathcal{W}(\bf{r}_1...\bf{r}_n)$ is defined in~\cite{Roux1999} as the potential of mean force of $n$ ions located in the channel at coordinates $ \bf{r}_1, ..., \bf{r}_n$.

Bearing in mind the definition (\ref{eq:config_integr}), equation (\ref{eq:grand_canonical}) for the probability of finding $ n $ ions in the filter (with $m=1$) reduces to the $ n $-ion binding factor given by equation (17) in~\cite{Roux1999}.

Equation (\ref{eq:config_integr}) serves as an operational definition of $\bar{\mu}^c +  \varepsilon\left( n ,n_f\right)$ in the molecular dynamics simulations.

\section{Conductivity of the filter in the linear response regime}\label{s:conduction}

In equilibrium, the electrochemical potentials $\mu_i$ are constant across the system for each species and the corresponding Nernst resting potentials ($\Delta\phi^{eq}_i = \phi^R-\phi^L$) are defined by the usual expression (cf.\ \cite{Roux1999})
\begin{eqnarray}\label{eq:nernst-pot}
\Delta\phi^{eq}_i = \frac{1}{q} \left[ \bar{\mu}^R_i- \bar{\mu}^L_i+ kT\ln \frac{c^R_i }{c^L_i} \right] =  \frac{kT}{q}\ln \frac{a^R_i }{a^L_i},
\end{eqnarray}
Here $a_i=\gamma_i c^s_i$ are activities with activity coefficients $\gamma_i$ defined by the equation $\bar{\mu}_i = kT \ln \gamma_i $  \cite{Friedman:86} and the concentrations $c^s_i$ are given by $c^s_i = x_i / V$.

Out of equilibrium, the current through the system is given \cite{Hille:01,Jackson2006} by the sum of the diffusion and conduction currents
\begin{eqnarray}\label{eq:total_current}
j_i = -q D_i \triangledown c_i - q c_i u_i \triangledown\phi.
\end{eqnarray}
where $u_i$ is the mobility.

Equation (\ref{eq:total_current}) can be rewritten following derivation of ~\cite{Landsberg1978,Landsberg1981} noting that, in equilibrium,
\begin{eqnarray}\label{eq:phi_grad}
q\triangledown\phi = \triangledown\mu_i - \triangledown\eta_i,
\end{eqnarray}
where $ \eta_i = kT \ln a_i$ and $ \mu_i$ are the chemical and electrochemical potentials respectively. On substituting (\ref{eq:phi_grad}) into the equation for the current, we have
\begin{eqnarray}\label{eq:modified_current}
	j_i = - c_i u_i \triangledown \mu_i -\left( q D_i \frac{\partial c_i}{\partial \eta_i} - c_i u_i \right) \triangledown \eta_i.
\end{eqnarray}
In equilibrium $ j_i =0 $ and $ \triangledown \mu_i = 0$, and one arrives~\cite{Landsberg1978,Landsberg1981} at the generalized Einstein relation~\cite{Ashcroft1976,Landsberg1978,Conway1983}
\begin{eqnarray}\label{eq:general_einstein}
\sigma_i = q^2 D \frac{\partial c_i}{\partial \eta_i}
\end{eqnarray}
between the ions' conductivity $\sigma_i = q c_i u_i $ and their diffusion constant $D$, and the concentration $c_i$ is defined as the mean number of ions in the filter divided by the filter volume.

Close to equilibrium the relation (\ref{eq:general_einstein}) still holds and from equation (\ref{eq:modified_current}) we obtain~\cite{Landsberg1978,Landsberg1981} (cf.\ \cite{Bisquert2008,Palenskis2013,Frenning2002,Bockris1973}) the current in the linear response regime as
\begin{eqnarray}\label{eq:current_Linear_response}
j_i = -\frac{\sigma_i}{q} \triangledown \mu_i, 
\end{eqnarray}
where the conductivity $\sigma_i$ is given by equation (\ref{eq:general_einstein}).

To calculate the filter conductivity, we note that the concentration of ions in the filter $c_i $ is proportional to the mean number of ions in the filter divided by the filter volume. The mean number of particles and the variance in the grand potential can be found as
\begin{equation}\label{eq:mean_N}
\left\langle n_i \right\rangle = -\left( \frac{ \partial\Omega}{\partial \tilde{\eta}_i}\right) _{T,V},
\end{equation}
\begin{equation}\label{eq:var_N}
\langle\left(\Delta n_i\right)^2\rangle   = kT\left( \frac{\partial\left\langle n_i \right\rangle}{\partial \tilde{\eta}_i}\right)_{T,V}.
\end{equation}
Here the chemical potential of the filter $\tilde{\eta}_i$ can be found as the difference between the free energies of the filter \cite{Kirkwood1935} containing either $(n_i+1)$ and or $n_i$ ions of a given type.

The theory also offers a way of estimating the coupling between the conducting ions observed in molecular-dynamics simulations~\cite{Liu2013a}. It can be estimated as follows
\begin{equation}\label{eq:covariance}
\langle n_i n_j\rangle - \left\langle n_i\right\rangle \left\langle n_j\right\rangle = kT\left( \frac{\partial\left\langle n_j \right\rangle}{\partial \tilde{\eta}_i}\right) _{T,V}.
\end{equation}

The generalized Einstein relation (\ref{eq:general_einstein}) plays a fundamental role in the conduction of systems with discrete numbers of particles and energy levels, including both quantum dots \cite{Beenakker:91,Alhassid:00}) and ion channels \cite{Kaufman:13a,Kaufman:12,Kaufman:15}.

This relationship is intuitively clear, because conduction occurs through a number of random steps. In each of these an extra ion enters and another ion leaves the filter. Therefore, the larger the number of such steps per second (i.e.\ the larger the fluctuations in ion number), the larger the conductivity of the filter becomes.

We will now apply the theory outlined above to the analysis of ions permeating the potassium channel. We will thereby demonstrate that it is the generalized Einstein relation and Coulomb blockade that underly its unusual~\cite{Eisenman61,Mullins1959,MacKinnon2001b,MacKinnon2003,Noskov2007a,Dixit2009a,Roux:11,Piasta:2011,Dixit2011a,Roux:2014} selectivity and conductivity properties.

\section{Application to the potassium filter}\label{s:potassium_filter}

Within the proposed formalism, the maxima in conductivity of the selectivity filter correspond to resonance-like peaks  \cite{Kaufman:13a,Kaufman:12,Kaufman:15} in the number of ionic fluctuations in the filter. In this section we will reveal how these maxima are related to the parameters of the potassium filter and to the Eisenman selectivity condition (\ref{eq:eisenman}).

To analyze permeation of the potassium filter, we assume that there are only two types of conducting ion in the bath solutions: Na$^+$ and K$^+$. The excess chemical potentials of these ions are $\bar{\mu}_{Na}$ and $\bar{\mu}_K$ in the baths, and $\bar{\mu}^c_{Na}$ and $\bar{\mu}^c_K$ in the filter.

Note that we do not discuss here the origin of the interaction at the binding site. Instead, the interaction is parametrized by the values of $\Delta\bar{\mu}_i$ and $\varepsilon(\{n_i\}, n_f)$, and the effect of these parameters on the filter permeation is then analyzed.

Note also that the analysis involve some additional approximations. For example, it is assumed that the binding sites have the same structure and that the channel pathway is a dielectric with constant dielectric permittivity $\epsilon_w$ \footnote{The latter assumption within our model affects the value of self-capacitance $C$. To avoid the limitations imposed by this assumption the value of $C$ can be evaluated experimentally using e.g.\ the equation for the charge fluctuations (\ref{eq:charge_fluctuations}) given in the appendix.}.

The values of the parameters $\Delta\bar{\mu}_K$ and  $\Delta\bar{\mu}_{Na}$ are chosen in the numerical examples below to guarantee that the free energy barrier in equation (\ref{eq:eisenman}) is reasonably close to earlier estimates~\cite{Noskov2007a,Dixit2011a,MacKinnon2001b}.

The total charge on the filter wall $qn_f$ should be viewed as an estimate of the charge facing the ions' pathway~\cite{MacKinnon2003c}. In these estimations we follow MacKinnon idea of charge balance~\cite{MacKinnon2003c} and the observation~\cite{MacKinnon2001b} that the number of K$^+$ ions in the filter fluctuates between 2 and 3. We take the corresponding effective charge to be around $ -2.5q$, i.e. approximately $-0.6$ per site, cf.~\cite{Allen1999a}.

We emphasize, however, that the mechanisms of permeation and selectivity described below do not depend on the exact values of the model parameters.

\subsection{States of the filter}\label{ss:states_filter}

As a first step in our analysis we find possible arrangements of the ions in the filter. We recall that only one ion at a time can bind to a given binding site in the single-file approximation (see Fig.~\ref{fig:channel}). I.e. each site can be in one of three states: empty (containing a water molecule) or filled with one potassium or with one sodium ion
\begin{equation}\label{eq:one_site_state}
	\left\lbrace 0,K^+,Na^+ \right\rbrace .
\end{equation}

In accordance with experimental observations, we allow the filter to hold at most three ions simultaneously. This assumption is consistent with the fact that the self-energy barrier becomes prohibitively large for the filter to hold more than three ions if the total negative charge on the channel wall is less than $3q$. In this case the full state space of the filter with indistinguishable binding sites is
\begin{eqnarray}\label{eq:full_space}
	&&\{ 000 \},~\{ K^+00 \},~\{ Na^+00 \},~\{ K^+K^+0 \},\nonumber\\
	&&\{ Na^+Na^+0 \},~\{ K^+Na^+0 \},~\{ K^+K^+K^+ \},\\
	&&\{ Na^+Na^+Na^+ \},~\{ Na^+Na^+K^+ \},~\{ K^+K^+Na^+ \}.\nonumber
\end{eqnarray}
The energy levels are associated with distinct configuration of ions in the filter. The shapes and positions of these levels depend on the properties of the filter, the concentration of conducting ions in the baths, and the applied voltage. We will now consider this dependence in more detail.

\subsection{Energy levels}\label{ss:Energy_levels_filter}

The free energy of the filter containing $n_K$ and $n_{Na}$ ions  is
\begin{eqnarray}\label{eq:G_Na_K}
&&G=-n_K \Delta\bar{\mu}_K -n_{Na} \Delta \bar{\mu}_{Na}-(n_K+n_{Na}) q\Delta\phi^s \nonumber\\
&&-kT\ln \frac{(x_K)^{n_K}}{n_K!} \frac{(x_{Na})^{n_{Na}}}{n_{Na}!}+\varepsilon(n_K,n_{Na},n_f),
\end{eqnarray}
where $s$ refers to either left ($s=L$) or right ($s=R$) bath, see Sec.~\ref{s:model_of_SFF}\ref{ss:energy}.

\begin{figure} [h!]
	\includegraphics[width=1.\linewidth]{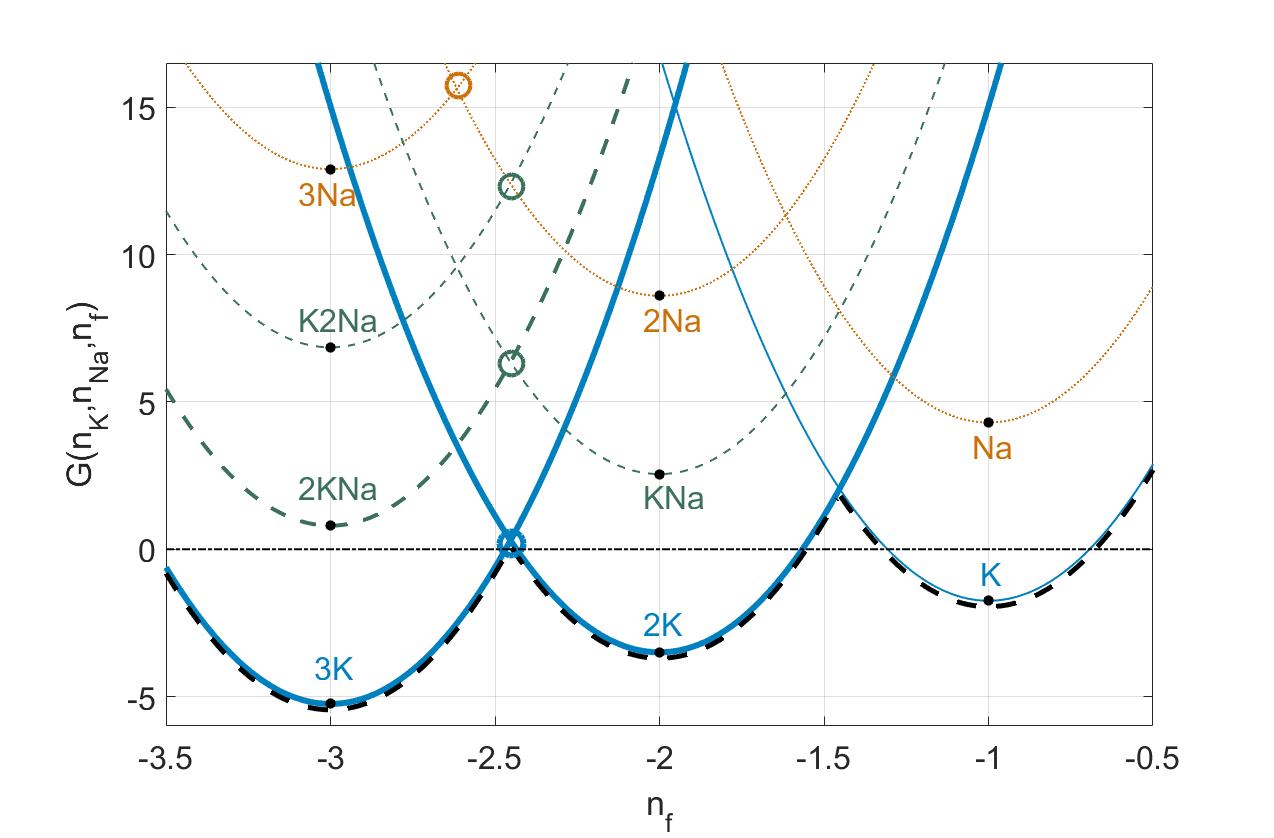}
	\vspace{-0.5cm}\caption{Levels of free energy for the following parameters: $\varDelta\bar{\mu}_K \approx -1.75$ kT; and $\varDelta\bar{\mu}_{Ns} \approx 4.3$ kT. Black dots show positions of the minima of the energy levels. Open circles show locations of the resonant conduction for $ n_f\approx-2.45$. The reference state is shown by horizontal black dashed-dotted line. $K$ energy levels are shown by blue solid lines and $Na$ levels are shown by yellow dotted lines. The mixed energy levels (filter filled with $K$ and $Na$) are shown by green dashed lines. \label{fig:free_energy_levels}}
\end{figure}

The self-energy barrier for the charges in the filter pathway is of the form (see Sec.~\ref{ss:ion_filter_interaction})
\begin{equation}\label{eq:electrostatic_energy_filter}
\varepsilon(n_K, n_{Na}, n_f) = \frac{q^2}{2C}\left(n_K+n_{Na}+n_f \right)^2 ,
\end{equation}
where $n_f$ is the fractional number of fixed negative charges on the filter wall. 

In total, there are ten different energy levels in the system, corresponding to the ten possible configurations of conducting ions given by (\ref{eq:full_space}). Nine of them are shown in Fig.~\ref{fig:free_energy_levels} as functions of the fixed charge on the wall $ n_f$ (and the tenth level corresponds to an unoccupied channel). The estimations were performed for a filter of length $L=12 \AA{}$, radius $R=1.5\AA{}$, and dielectric constant $\epsilon_w = 80$. The resultant value of $\frac{q^2}{2C} \approx 18.5 kT$, $\triangle\phi=0$, and values of $\Delta\bar{\mu}_K$ and  $\Delta\bar{\mu}_{Na}$ are given in the figure caption.

From Fig.~\ref{fig:free_energy_levels} it is evident that there are four sets of parabolic energy levels. The lowest ones (blue solid lines) correspond to the filter being filled with one, two, or three K$^+$ ions, respectively. The highest energy levels (orange dotted lines) correspond to the system states with Na$^+$, 2Na$^+$, and 3Na$^+$. The mixed states of the filter filed with K$^+$ and Na$^+$ ions are of intermediate energy as shown by green dashed lines.

We emphasise that the levels are quadratic functions of $n_f$ in accordance with equation (\ref{eq:electrostatic_energy_filter}). The curvature at their minima is determined by the value of $U_c$.

The energy level minima shown by the black dots correspond to the condition
\[ n_K + n_{Na} + n_f = 0.\]
If we neglect the contributions to $G$ coming from the electrochemical potential, all the minima will then be at zero energy corresponding to the reference state shown in the figure by the horizontal black dashed-dotted line. In this case we recover the situation analyzed in our earlier work~\cite{Kaufman:12,Kaufman:13a,Kaufman:15} with symmetric solutions in the absence of dehydration.

In general, the location of the minimum in each case is determined by the value of the electrochemical potential given by first four terms in equation (\ref{eq:G_Na_K}).

The position of the levels for various types of ion provides an insight into the selectivity filter's population, selectivity, and conductivity as will now be discussed in more detail.

\subsection{Filter occupancy and conductivity}\label{ss:Filter_conduction}

To analyze the channel occupancy and $\langle \Delta n^2_i \rangle $, we use the following equations which are equivalent to the definitions (\ref{eq:mean_N}) and (\ref{eq:var_N})
\begin{eqnarray}\label{eq:mean_and_variance}
&& \langle n_i \rangle  = \sum_{\{n_i\}} n_i P(n_i,n_f)\\
&& \langle n^2_i \rangle = \sum_{\{n_i\}} n^2_i P(n_i,n_f), \quad  \langle \Delta n^2_i \rangle = \langle n^2_i \rangle -\langle n_i \rangle^2. \nonumber
\end{eqnarray}
The binding probabilities $P(n_i,n_f)$ in equations (\ref{eq:mean_and_variance}) are given by (\ref{eq:grand_canonical}), and the summation is carried out over all configurations of the ions in the filter.

\begin{figure}[h!]
	\includegraphics[width=0.95\linewidth]{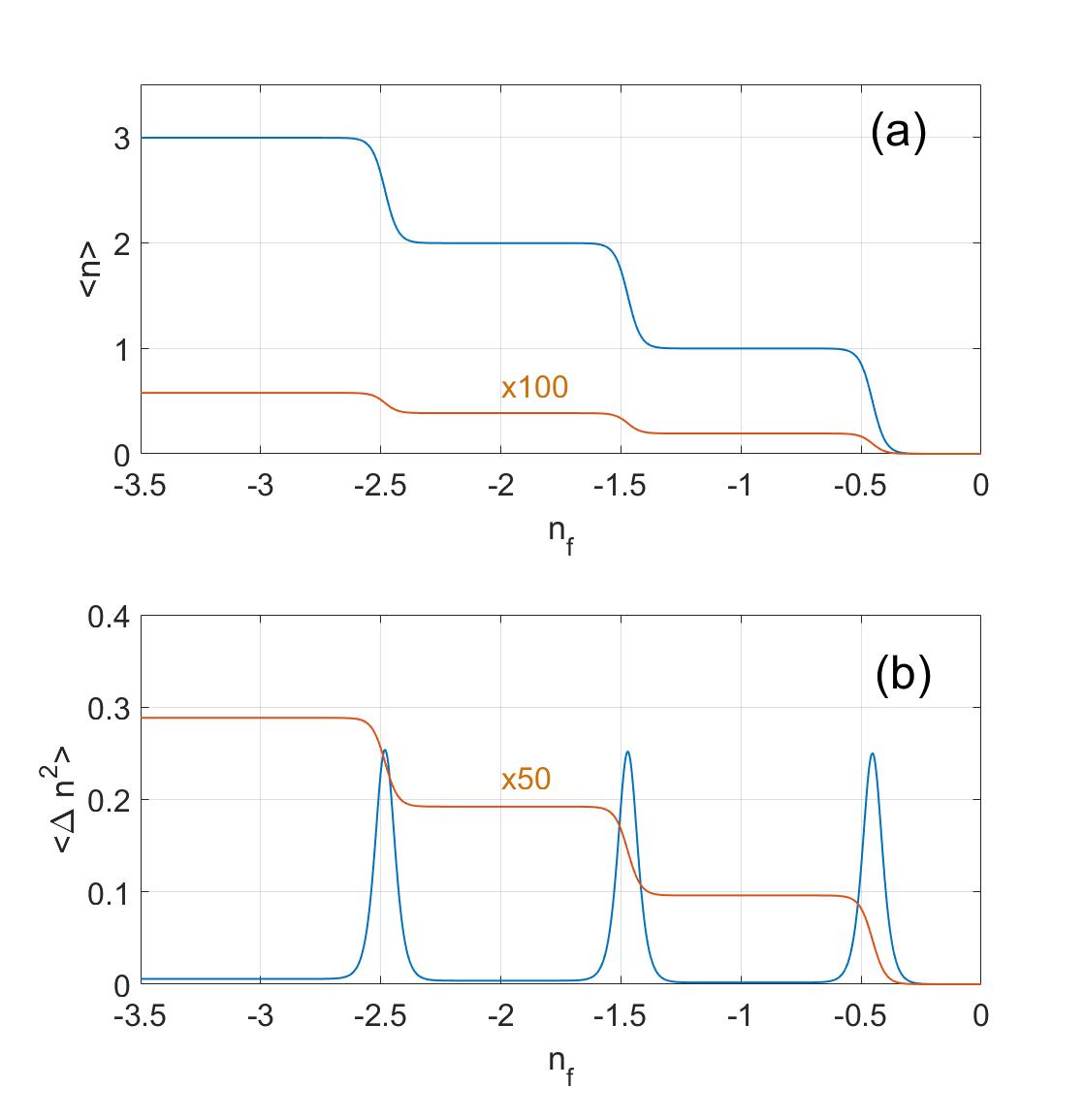}
	\caption{(a) Mean and (b) variance for K$^+$ (blue lines) and $Na^+$ (orange lines) ions in the channel. The mean value for $ Na^+$ ions in figure (a) is multiplied by 100. the variance for $ Na^+$ ions in figure (b) is multiplied by 50. }
	\label{fig:staircase1}
\end{figure}

The resultant occupancies and their fluctuations are shown in Fig.~\ref{fig:staircase1} (a) for the same model parameters as in Fig.~\ref{fig:free_energy_levels}. It can be seen that the occupancy of K$^+$ ions has the staircase-shape familiar from research on quantum dots~\cite{Glazman1999} and our earlier work on valence selectivity~\cite{Kaufman:13a,Kaufman:12,Kaufman:15}. At each step the average number of K$^+$ ions in the filter increases by one, starting from zero ions at $n_f=0$ and ending with three K$^+$ ions at $n_f=-3$.

The occupancy by sodium ions remains smaller than $0.01$ for all values of $ n_f$ and physiological values of the other model parameters.

The fluctuations in the number of K$^+$ ions in the filter exhibit a sharp peak at the location of each step, whereas fluctuations in the number of sodium ions in the channel are very small for all values of $n_f$, as shown in Fig.\ ~\ref{fig:staircase1}(b).

According to linear response theory the current through the filter is proportional to the $\langle \Delta n^2_K \rangle $ indicating that K$^+$ conductivity will be sharply peaked at the transition point, approaching the diffusion limit: see equation (\ref{eq:general_einstein}).

This result is intuitively clear, because these points correspond to degeneracies in the system (cf.\ \cite{Glazman1999}), when adding or removing the third K$^+$ ion to or from the filter does not cost any energy. It was pointed out in our earlier work~\cite{Kaufman:13a,Kaufman:12,Kaufman:15} that ionic transfer through the filter at the transition points corresponds to the ``knock-on'' mechanism of conduction within Coulomb blockade theory.

On the other hand, when the filter fits ``perfectly'' (i.e.\ its energy has a minimum) for an integer number of K$^+$ ions, its conduction is blocked. This result of equilibrium statistical theory provides an insight into the earlier argument~\cite{Roux:2014} that the ``snug-fit'' model cannot describe ionic conduction through the filter at nearly the diffusion speed.

It follows from this discussion that the condition for maximum K$^+$ conductivity is
\begin{eqnarray}\label{eq:cond_max_cond}
G(2K^+,n_f) = G(3K^+,n_f).
\end{eqnarray}
If we require barrier-less conduction of the K$^+$ ions, the transition point must also satisfy an additional condition
\begin{eqnarray}\label{eq:cond_barrier_less}
G(2K^+,n_f) = G(3K^+,n_f) \approx 0.
\end{eqnarray}
We note that this second condition for a given set of parameters can be only satisfied for the $2K^+ \to 3K^+$ transition, in agreement with experimental observation~\cite{MacKinnon2001b}.

The ability of nano-filters to ``resonantly'' conduct ions for a specific value of negative charge on the  channel wall, and to block an ion's  passage through the filter at a different value of the fixed charge, is the essence of the ionic Coulomb blockade theory of biological channels~\cite{Kaufman:13a,Kaufman:12,Kaufman:15} and of artificial nanopores as originally envisaged in \cite{Krems:13}. The same property underlies the  ``band structure'' of the ionic current through channels discovered in our earlier work~\cite{Kaufman:13a,Kaufman:12}.

To understand how these nano-filters can select between ions of the same valence for a given fixed charge $n_f = -2.45$ on the wall, we note that there are three transition points corresponding to the ``knock-on'' mechanism of  conductivity at value of $n_f$, as shown in Fig.~\ref{fig:free_energy_levels}. Only the lowest-energy transition point corresponds to the pure K$^+$ conduction with $G(2$K$^+,n_f) = G(3$K$^+,n_f) \approx 0$.

All other transition points correspond to mixed conductivity. The first one, at the the transition point KNa$\rightarrow 2$KNa, corresponds to conduction involving two K$^+$ ions one Na$^+$ ion.  The second one, at the the transition point Na$\rightarrow$K2Na, corresponds to conduction involving one K$^+$ ion two Na$^+$ ions.

Fig.~\ref{fig:free_energy_levels} shows that the transition points corresponding to mixed conductivity have high potential barriers, and that their height increases with increasing numbers of participating sodium ions. Thus the first and second mixed conductivity points have barriers of $\Delta G\approx 6kT$ and $\Delta G\approx 12kT$, respectively.

It is evident that pure sodium conduction would have to overcome the highest free energy barrier of all ($\approx 16kT$) and also that it is characterised by a different value of $n_f \simeq -2.61$.

We conclude, therefore, that the conductivity of sodium through the filter would be impeded by prohibitively large free energy barriers, effectively blocking its conduction. We will now derive the equation for this barrier under the condition that the filter is tuned to conduct K$^+$ ions at the maximum rate.

\subsection{Generalized Eisenman selectivity of the filter}\label{ss:Filter_selectivity}

It was shown in the previous subsection that the KcsA selectivity filter can conduct K$^+$ ions at a rate close to that of free diffusion provided that two conditions fulfilled:
\[G(n_K,n^*_f) = G(n_K+1,n^*_f)~ \text{ and }  ~G(n_K,n^*_f) \approx 0.\]
These two conditions define the optimal $n^*_f$ and optimal excess chemical potential $\Delta\mu_K^*$ in terms of the characteristic electrostatic self-energy $U_c$, applied voltage, and bulk concentrations in equilibrium. The optimal values of $n^*_f$ and $\Delta\mu_K^*$ are obtained by solving simultaneously the following equations
\begin{eqnarray}\label{eq:K_resonance}
&& n^*_f = -(n_K+1/2)+\frac{C}{q^2}\Delta\mu^*_K, \\
&& \Delta\mu^*_K \approx \frac{1}{n_K}\varepsilon(n_K,0,n^*_f).
\end{eqnarray}
Here $\Delta\mu_K = \Delta\bar{\mu}_K + q\Delta\phi^s + kT\ln \left( x_K/n_K!\right)$.

Substituting these values into the free energy $ G(n_K,n_{Na},n_f^*)$ for a filter that contains one additional Na$^+$ ion gives the following barrier for the latter ion to enter the channel
\begin{eqnarray}\label{eq:Na_barrier}
&& \Delta G_{Na} = (\bar{\mu}^c_{Na}- \bar{\mu}_{Na})\\
&&\qquad - (\bar{\mu}^c_K- \bar{\mu}_K) + kT\ln\frac{c_K}{n_K!c_{Na}}. \nonumber
\end{eqnarray}
Thus we can see that the Eisenman selectivity relation (\ref{eq:eisenman}) follows directly from the equilibrium statistical theory of the filter under conditions of fast diffusion-limited conduction of K$^+$ ions.

We note the presence of an additional term proportional to the logarithm of the ion concentration ratio in the bulk. To emphasize the influence of concentration on the selectivity of the filter, we will call the relation (\ref{eq:Na_barrier}) the {\it generalized Eisenman relation}.

It is also interesting to note that these results shed new light on the relationship between the filter selectivity and the large fluctuations of the channel walls observed experimentally. Indeed, fluctuations of the filter walls are already included in the energy levels and selectivity conditions of the system. These conditions are obtained by averaging the system response on a timescale much larger than the characteristic timescale of fluctuations, but much smaller than an ion's time of passage through the system.

We therefore conclude that, within statistical theory of the selectivity filter, the main effect of wall fluctuations is an increase of the diffusion coefficient $D$ in the equation for the current (\ref{eq:current_Linear_response}) and therefore an increase in the conductivity of the system.

\section{Conclusions}\label{s:conclusions}

In summary, we have introduced and developed a statistical mechanical model of conductivity and of selectivity between ions of the same valence in biological selectivity filters. The theory consists of two main parts: a grand canonical ensemble for the ionic distribution in the filter coupled to two baths with mixture of conducting ions of arbitrary concentration; and the equation for the current through the filter in the linear response regime with the conductivity being given by the generalized Einstein relation.

We formulated the conditions required for high diffusion-limited barrier-less conductivity in such filters and showed that, in the presence of dehydration, these conditions are valid only for one of the conducting ion species.

We then applied this theory to the analysis of the selectivity and conductivity of the KcsA filter and proposed a resolution of the long-discussed paradox of its high selectivity being combined with high conductivity. The paradox is resolved by showing that the Eisenman relation for filter selectivity follows directly from the condition for fast barrier-less conduction of K$^+$ ions.

The results obtained also illuminate the long-discussed relationship between the ``snug-fit'' model and experimentally-observed large fluctuations of the channel wall. We demonstrated that a filter that is  ``perfectly'' fitted to accommodate an integer number of ions does not conduct. In sharp contradistinction, high conductivity corresponds to the situation when the filter is tuned to have an equal probability of containing either 2 or 3 ions. In our model, the large wall fluctuations result solely in an increase of the effective diffusion coefficient (and thus conductivity) of the filter, without affecting its selectivity.

The proposed theory is also applicable to the analysis of the current though artificial nano-pores ~\cite{Jain2015,Feng2016}, for which the corresponding analysis will be a subject of future research.

The results obtained can be extended to encompass filters with distinguishable binding sites and mixed conduction by mono-  and divalent ions.

\vspace*{0.5cm}

\appendix

\section{Charge fluctuations }\label{a:Charge_fluctuations}

It can be seen from equations (\ref{eq:general_einstein}), (\ref{eq:current_Linear_response}), and (\ref{eq:var_N}) that the current though the channel is proportional to the variance of the number of conducting ions in the filter. This is a fundamental property of systems where the discrete nature of the conducting  particles becomes apparent.

The conductivity $\sigma_i$ in this system is given by the Kubo formula \cite{Kubo:66}
\begin{equation}\label{eq:Kubo}
\sigma_i = \frac{1}{kT}\int_0^\infty d\tau  \langle  j_i(t+\tau) j_i(t) \rangle,
\end{equation}
where the instantaneous current of $i$-th ions with velocities $v_{i,k}(t)$ is
\[j_i(t)=\sum_{k=1}^{n_i} q v_{i,k}(t).\]
It can further be shown (by introducing $Q(t) = \int_{0}^{t}j_i(\tau)d\tau$ and substituting into (\ref{eq:Kubo}), \cite{Helfand1960,Chandler1987,Liu2013a,Frenkel1996}, cf ~\cite{Roux:04}) that the relationship between conductance and mean-square displacement of the charge transferred through the system takes the form
\begin{equation}\label{eq:conductivity_charge_fluct}
\sigma_i = \frac{1}{2kT} \frac{d \langle  Q^2(t)\rangle}{d t}.
\end{equation}
Indeed, it is well known that, for a small conductor in equilibrium with a large reservoir of heat and particles, the total charge fluctuations depend on the associated capacitance ~\cite{Landsberg2014} as
\begin{eqnarray}\label{eq:charge_fluctuations}
\langle Q^2 \rangle = q^2 \langle \left( \Delta n_i\right) ^2 \rangle = C kT.
\end{eqnarray}
The corresponding autocorrelation function $\langle  Q^2(t)\rangle \propto \langle Q^2 \rangle \exp\left( -t/RC\right) $, can be viewed as a consequence of the Onsager regression hypothesis~\cite{Chandler1987} applied to the parallel $RC$-circuit model of the channel (cf.\ a different proof given in ~\cite{Landsberg2014}). We conclude that the channel conductivity is proportional to the charge fluctuations and thus to particle number fluctuations.

\begin{acknowledgments}
We acknowledge valuable discussions with Stephen Roberts, Olena Fedorenko, Carlo Guardiani, Igor Khovanov, and Aneta Stefanovska. The research has been supported by the Engineering and Physical Sciences Research Council UK (grant No.\  EP/M015831/1).
\end{acknowledgments}


\bibliography{ionchannels}
\end{document}